\documentclass[review]{elsarticle}
\usepackage{amssymb}
\usepackage{lineno,hyperref}
\modulolinenumbers[5]

\journal{Physics of the Dark Universe}
\def\beq{\begin{equation}}
\newcommand{\eeq}[1]{\label{#1}\end{equation}}
\def\bea{\begin{eqnarray}}
\newcommand{\eea}[1]{\label{#1}\end{eqnarray}}
\newcommand{\be}{\begin{eqnarray}}
\newcommand{\ee}{\end{eqnarray}}
\begin{document}
\begin{frontmatter}
%
%
%
%

\title{
\begin{flushright}
{\small CERN-PH-TH/2015-181 \\}
\end{flushright} 
$\phantom{a}$ \\
Pre--Inflationary Relics in the CMB ?}
%
%


\author[mymainaddressAG,mysecondaryaddressAG]{A.Gruppuso}
\ead{gruppuso@iasfbo.inaf.it}

\author[mymainaddressNK]{N.~Kitazawa}
\ead{kitazawa@phys.se.tmu.ac.jp}

\author[mymainaddressNM,mymainaddressAG]{N.~Mandolesi\corref{mycorrespondingauthor}}
\cortext[mycorrespondingauthor]{Corresponding author}
\ead{mandolesi@iasfbo.inaf.it}

\author[mymainaddressNM]{P.~Natoli}
\ead{paolo.natoli@unife.it}

\author[mymainaddressAS,mysecondaryaddressAS]{A.~Sagnotti}
\ead{sagnotti@sns.it}

\address[mymainaddressAG]{INAF-IASF Bologna,
Istituto di Astrofisica Spaziale e Fisica Cosmica di Bologna,
Istituto Nazionale di Astrofisica,
via Gobetti 101, I-40129 Bologna, Italy}
\address[mysecondaryaddressAG]{INFN, Sezione di Bologna,
Via Irnerio 46, I-40126 Bologna, Italy}
\address[mymainaddressNK]{Department of Physics, Tokyo Metropolitan University,
Hachioji, Tokyo 192-0397, Japan}
\address[mymainaddressNM]{Dipartimento di Fisica e Scienze della Terra and INFN, Universit\`a degli Studi di Ferrara, Via Saragat 1, I-44100 Ferrara, Italy,
INAF-IASF Bologna}
\address[mymainaddressAS]{Department of Physics, CERN Theory Division,
CH - 1211 Geneva 23, Switzerland}
\address[mysecondaryaddressAS]{Scuola Normale Superiore and INFN,
Piazza dei Cavalieri 7\
I-56126 Pisa, Italy}

%
%
%
%
%
%
\begin{abstract}
\noindent String Theory and Supergravity allow, in principle, to follow the transition of the inflaton from pre-inflationary fast roll to slow roll. This introduces an infrared depression in the primordial power spectrum that might have left an imprint in the CMB anisotropy, if it occurred at accessible wavelengths. We model the effect extending $\Lambda$CDM with a scale $\Delta$ related to the infrared depression and explore the constraints allowed by {\sc Planck} 2015 data, employing also more conservative, wider Galactic masks in the low resolution CMB likelihood. In an extended mask with $f_{sky}=39\%$, we thus find $\Delta = (0.351 \pm 0.114) \times 10^{-3} \, \mbox{Mpc}^{-1}$, at $99.4\%$ confidence level, to be compared with a nearby value at $88.5\%$ with the standard $f_{sky}=94\%$ mask. With about 64 $e$--folds of inflation, these values for $\Delta$ would translate into primordial energy scales ${\cal O}(10^{14})$ GeV.
\end{abstract}

\end{frontmatter}



%
%
%
%
\section{Introduction}

As a unified framework for all interactions, String Theory \cite{stringtheory} ought to provide some insights into the Early Universe, but attempts to extract concrete information have foundered on our incomplete grasp of its key principles. On the other hand, the low--energy Supergravity \cite{sugra} leaves aside higher--derivative stringy corrections, which casts a shadow of doubt on the resulting dynamics. Probably also as a result of these facts, the analysis of inflation \cite{inflation} has been largely confined to its steady state.

Slow--roll models with a single scalar field yield the power spectra of scalar perturbations \cite{cm}
\beq
{\cal P}(k) \ = \ A \ \left(k/k_0\right)^{n_s-1} \ ,
\eeq{tilt}
where the amplitude $A$ reflects typical energy scales during inflation and $k_0$ is a pivot scale\footnote{We set $k_0=0.05 \,{\rm Mpc}^{-1}$, checking however that this standard choice has no impact on the ensuing analysis.}.
{\sc Planck} recently obtained the result $n_s~=~0.968\,\pm \,0.006$ for the spectral index \cite{Planck:2015xua}, so that this peculiar behavior finds indeed a place in the CMB. There are, however, some intriguing discrepancies with the resulting $\Lambda$CDM picture, including an apparent lack of power at large angular scales, with a sizable quadrupole depression.

The discrepancies appear in the first few
angular power spectrum coefficients $C_\ell$, which then converge to the $\Lambda$CDM expectations within a decade or so. Theory
associates these low--$\ell$ values with the earliest accessible epochs of
inflation, so any
departure from $\Lambda$CDM would be of utmost interest.  There is, of
course, the issue of ``cosmic variance,'' since we can detect only a single
realization of the CMB anisotropy pattern.  Still, the discrepancies have surfaced in independent experiments, so
that explaining them in terms of systematics or unresolved foregrounds
would be contrived.  Hence, in this paper, we propose to take low--$\ell$
anomalies seriously, combining the relevant {\sc Planck} data with clues from String Theory and Supergravity.  This approach parallels the analysis in \cite{planck_XX}, but
is based on a different philosophy.

It is tempting, if not fully justified within Supergravity alone, to explore how slow--roll was originally attained. We shall insist on models with a single scalar field $\phi$, starting from a Bunch--Davies vacuum in a background
\beq
    ds^2 \ = \ e^{\,{2} \,{\cal A}(\eta)} \left(\, -\, d\eta^2 \, + \, \mathrm{d}\mathbf{x}\cdot \mathrm{d}\mathbf{x} \right) \ ,
\eeq{metric}
where the conformal time $\eta$ is conventionally set to zero at the end of inflation, while putting some emphasis on the approach to slow--roll.
${\cal A}(\eta)$ and $\phi(\eta)$ determine \cite{brand_mukh}
\beq
W_s(\eta) \ = \ \frac{1}{z} \ \frac{d^2 z}{d \eta^2}\ , \quad {\rm where} \quad z(\eta) \ = \ e^{\,{\cal A}} \ \frac{d \phi}{d {\cal A}} \ ,
\eeq{Ws}
and thus the Mukhanov--Sasaki equation
\beq
\frac{d^2 v_k(\eta)}{d\eta^2} \ + \ \left[ k^2 \ - \ W_s(\eta) \right] v_k(\eta) \ = \ 0 \ .
\eeq{ms}
The power spectrum of scalar perturbations that builds up after many $e$--folds of inflation is then
\beq
{\cal P}(k) \ = \ \frac{k^3}{2 \pi^2} \ \lim_{\eta \to 0^-} \ \left| \frac{v_k(\eta)}{z(\eta)} \right|^2 \ .
\eeq{power}
Close to an initial singularity, set here at $\eta=-\eta_0$, one can show that
\beq
W_s(\eta) \ \simeq \ - \ \frac{1}{4 \ \left( \eta \ + \ \eta_0 \right)^2} \ ,
\eeq{initial}
while after several $e$--folds
\beq
W_s(\eta) \simeq \frac{\nu^2 - \frac{1}{4}}{\eta^2} \ , \qquad \nu = 2 \ - \ \frac{n_s}{2} \ .
\eeq{inflation}
When exploring the onset of inflation one is thus confronted with \emph{Mukhanov--Sasaki potentials that cross the $\eta$ axis}. They bring along \emph{an infra-red depression sized by the measure factor} in (\ref{power}), so that ${\cal P}(k) \sim k^3$, but the approach to the profile (\ref{tilt}) is not universal \cite{dkps,ks}. For instance, subtracting from the Mukhanov--Sasaki potential (\ref{inflation}) a positive quantity $\Delta^2$ makes it cross the negative $\eta$--axis but turns $k^2$ into $k^2+\Delta^2$, and thus eq.~(\ref{tilt}) into the exact result
\beq
{\cal P}(k) \ = \  \frac{ A \, \left(k/k_0\right)^3}{\left[\left(k/k_0\right)^2 + \left(\Delta/k_0\right)^2\right]^{\nu}} \ ,
\eeq{power_cutoff}
which brings in the new scale $\Delta$.
\begin{figure}[ht]
\centering
\begin{tabular}{cc}
\includegraphics[width=30mm]{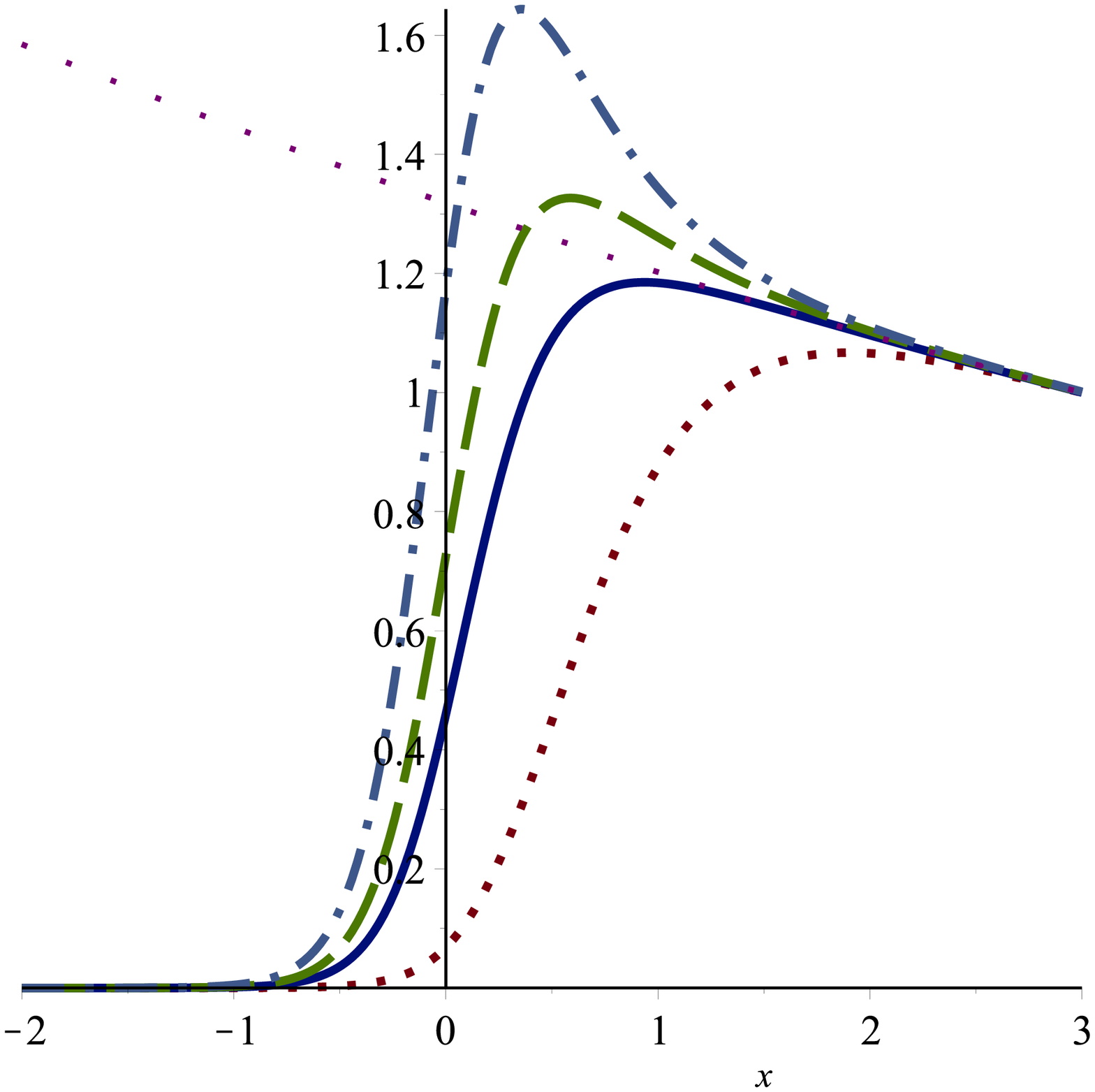} & \quad
\includegraphics[width=30mm]{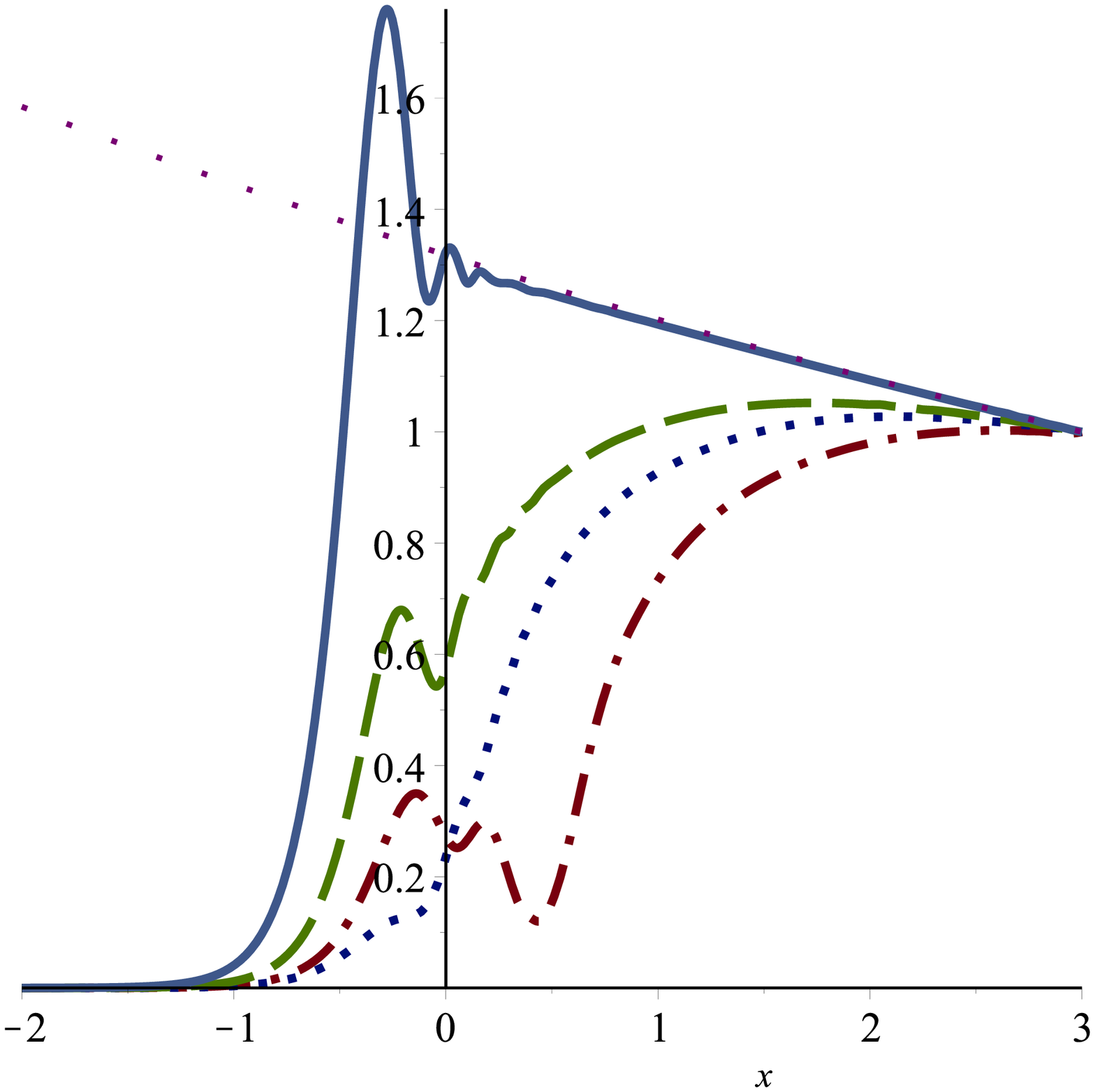} \\
\end{tabular}
\caption{\small  Some power spectra in arbitrary units \emph{vs} $x=\log_{10}(k/k_0)$, compared with the light dotted curves obtained from eq.~(\ref{tilt}). \emph{Left:} analytic power spectra from eq.~(\ref{coulomb_power}) for $\gamma=-1$ (dotted), $0$ (continuous), $0.2$ (dashed), $0.4$ (dashed-dotted). \emph{Right:} four types of spectra from BSB, where the scalar bounces against a steep exponential potential before attaining slow--roll. An initial condition, $\varphi_0$, gauges the bounce, and mild bounces (continuous) recover the spectra of \cite{destri}. }
\label{fig:compared_spectra}
\end{figure}

More generally, for the Coulomb--like potentials
\beq
{ W_s \ = \ \frac{\nu^2 - \frac{1}{4}}{\eta^2} \left[ c \left( 1
+ \frac{\eta}{\eta_0}\right) + (1-c) \left( 1
+ \frac{\eta}{\eta_0}\right)^2 \right]}
\eeq{coulomb_pot}
the Mukhanov--Sasaki problem admits the family of exact solutions \cite{dkps}
\begin{eqnarray}
&&{\cal P}(k) = \frac{ A \, \left(k/k_0\right)^3 {\cal C}(k)}{\left[\left(k/k_0\right)^2 + \left(\Delta/k_0\right)^2\right]^{\nu}} \label{coulomb_power} \, , \\
&& {\cal C}(k) =  \frac{\Gamma \left(\nu + \frac{1}{2}\right)^2\, e^{\pi \, B(k)}}{\left|\Gamma \left(\nu + \frac{1}{2} + i B(k)\right)\right|^2} \ ,
 \ \Delta^2 = \frac{(c-1)\left( \nu^2 - \frac{1}{4}\right)}{\eta_0^2} \nonumber \\
&&B(k) = \frac{\gamma}{\sqrt{\left(\frac{k}{k_0}\right)^2 + \left(\frac{\Delta}{k_0}\right)^2}} \ , \  \gamma =  \frac{\left(\frac{c}{2} - 1\right)\left(\nu^2 - \frac{1}{4}\right)}{k_0\,\eta_0} \,  \nonumber\ .
\end{eqnarray}

For $1< c < 2$, these power spectra are along the lines of the $c=2$ case of eq.~(\ref{power_cutoff}), but for $c>2$ a caricature pre--inflationary peak builds up. It lies next to the almost scale invariant profile, as in \cite{destri}, since $\Delta$ enters both factors in eqs.~(\ref{coulomb_power}). On the other hand (see fig.~\ref{fig:compared_spectra}), in the orientifold vacua \cite{orientifolds} of String Theory with ``Brane Supersymmetry Breaking'' (BSB) \cite{bsb,climbing} pre--inflationary peaks can lie well apart from the limiting profile, an option that appears favored by low CMB multipoles \cite{ks}.

The models of eqs.~(\ref{power_cutoff}) and (\ref{coulomb_power}) provide a convenient point of departure from $\Lambda$CDM. We shall first analyze the new scale $\Delta$, and then we shall also include $\gamma$ to take a first look at pre--inflationary features. Our approach differs from the previous work in \cite{destri} that also addressed the full spectrum in three respects. The first is the emphasis on eq.~(\ref{power_cutoff}), which is motivated by the Mukhanov--Sasaki equation and, as we have seen, by the sign change of $W_s$ that accompanies the approach to slow-roll. The others are the use of recent {\sc Planck} 2015 data and, as we about to explain, the use of different Galactic masks.

\section{Data Set}

\begin{figure}[ht]
\centering
\begin{tabular}{c}
\includegraphics[width=55mm]{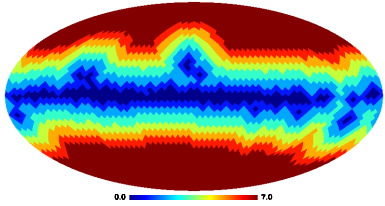}
\end{tabular}
\caption{\small Temperature masks adopted in the low--$\ell$ {\sc Planck} likelihood. The color coding 0 identifies the standard mask, while combinations identify its extensions. Thus, 0 and 1 identify the mask extended by $6^\circ$,
0,1 and 2 the mask extended by $12^\circ$, 0,1,2 and 3 the mask extended by $18^\circ$, 0,1,2,3 and 4 the mask extended by $24^\circ$. Finally, 0,1,2,3,4 and 5 and
0,1,2,3,4,5 and 6 identify the masks extended by $30^\circ$ and $36^\circ$.}
\label{fig:masks}
\end{figure}

We used the recently released {\sc Planck} likelihood module \cite{PlanckLike}, considering the CMB temperature (TT), low--$\ell$ polarization (lowP) and lensing likelihoods.
We sampled over the six standard $\Lambda$CDM cosmological parameters
($\theta_{MC}$, $\Omega_b h^2$, $\Omega_c h^2$, $n_s$, $\tau$ and $\ln (10^{10} \, A_s)$) \cite{Planck:2015xua} and over $\Delta$, which models the low--$\ell$ depression via eq.~(\ref{power_cutoff}). We also sampled over the instrumental and foreground parameters that appear in the {\sc Planck} likelihood: we do not report on them here for the sake of brevity, but we did verify that their posteriors did not depart from the $\Lambda$CDM case.
We also implemented a conservative modification of the low $\ell$ part of the {\sc Planck} likelihood code, allowing for low--$\ell$ larger temperature masks.
The masks are {\it blindly} built, extending the edges of the standard temperature mask by fringes of widths $6^\circ$, $12^\circ$, $18^\circ$, $24^\circ$, $30^\circ$ and $36^\circ$, as shown in fig.~\ref{fig:masks}.
They reduce the allowed sky fraction, $f_\mathrm{sky}$, for $\ell \le 29$ from 94\% to 84\%, 71\%, 59\%, 49\%, 39\% and 31\%, and are applied to a foreground reduced CMB map based on the {\tt Commander} algorithm \cite{commander}, as in the original {\sc Planck} release.
Cosmological parameters should not depend on the specific part of the sky that is analyzed if the CMB pattern is isotropically distributed.
The fact that several low--$\ell$ CMB anomalies are enhanced
when portions of the sky close to the Galactic plane are excluded (see \emph{e.g.}~\cite{Copi:2008hw,Gruppuso:2013xba,Gruppuso:2013dba}) was a key motivation to test the stability of our results against Galactic masking.
The most interesting cases ($f_{sky}=94\%$, $f_{sky}=39\%$) were analyzed considering also {\sc Planck} polarization data at high $\ell$, thus taking into account TT, TE and EE information \cite{PlanckLike}. Unless otherwise specified, however, the results will be understood to contain only TT at high $\ell$.

\section{Results}

\begin{figure}[ht]
\centering
\begin{tabular}{cc}
\includegraphics[width=30mm]{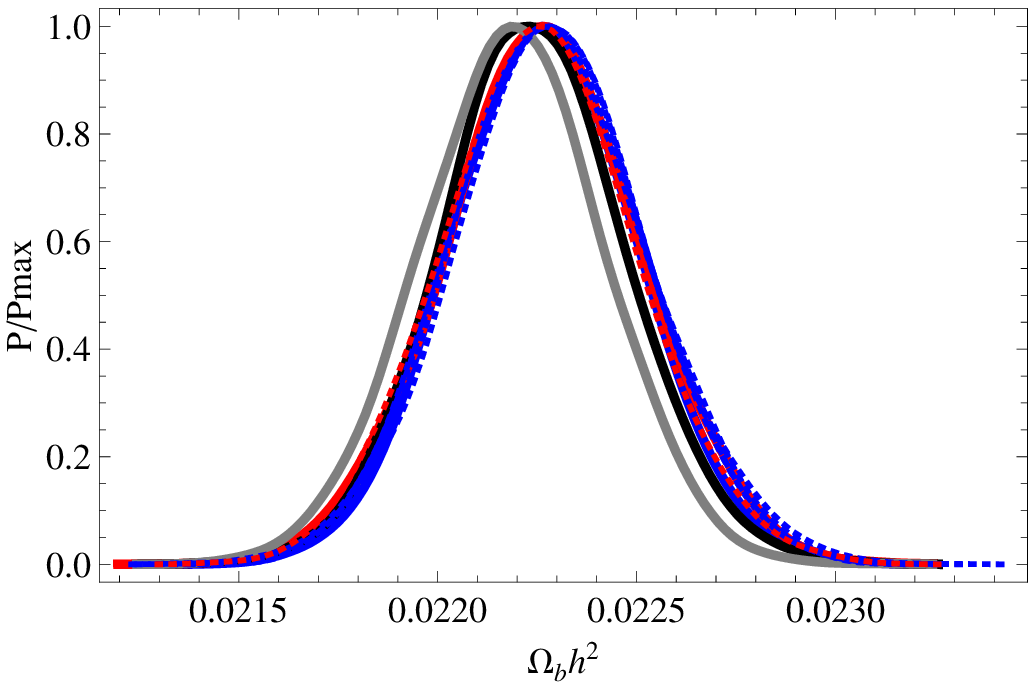} & \qquad
\includegraphics[width=30mm]{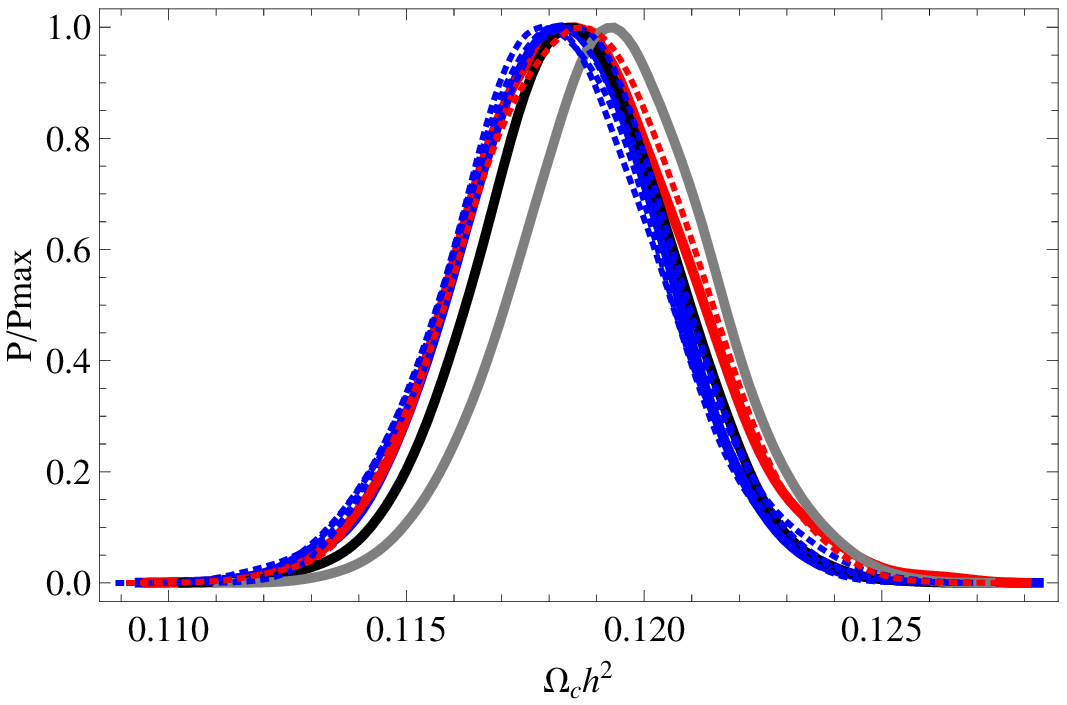} \\
\includegraphics[width=30mm]{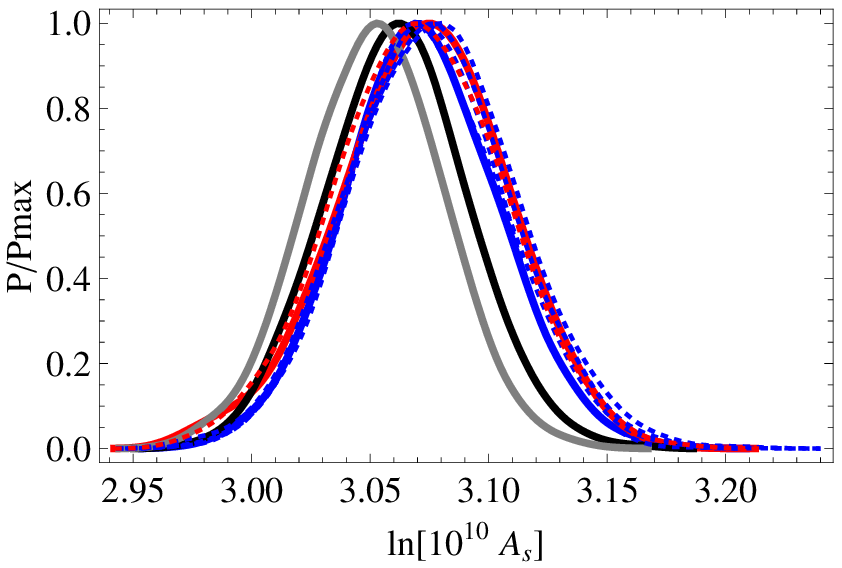} & \qquad
\includegraphics[width=30mm]{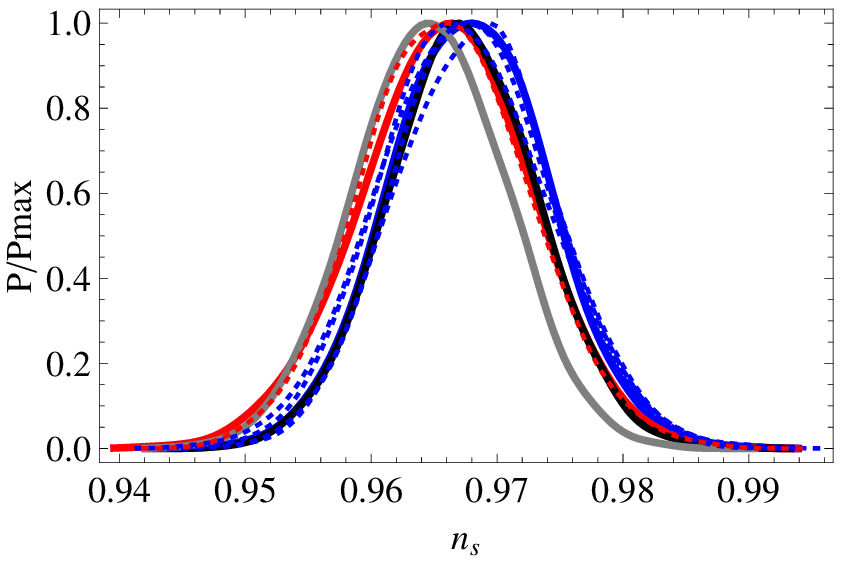} \\
\includegraphics[width=30mm]{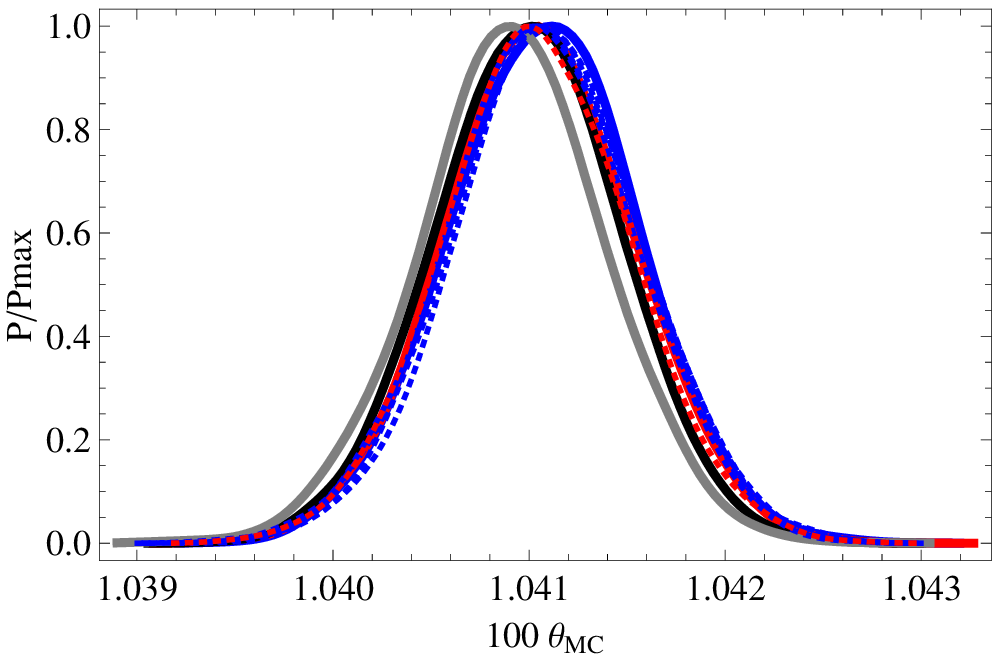} & \qquad
\includegraphics[width=30mm]{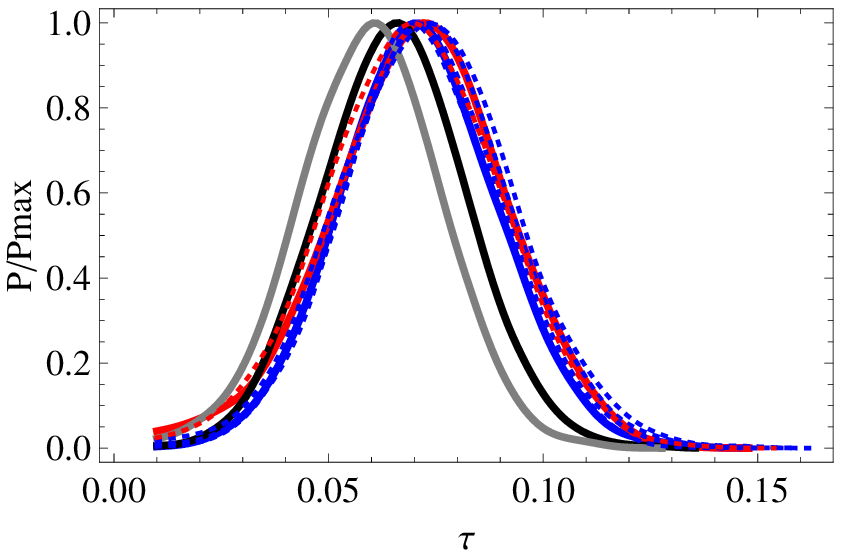} \\
\end{tabular}
\caption{\small Posteriors for the six standard $\Lambda$CDM cosmological parameters. The black and gray curves are determined by eq.~(\ref{tilt}), and refer to the standard 94\% mask released by {\sc Planck} (black) and to a $+30^\circ$ extended mask (gray).
All other curves rest on the modified power law of eq.~(\ref{power_cutoff}): solid blue for the 94\% mask, thick red for a $+30^\circ$ extension, dotted blue for the intermediate masks $+6^\circ$, $+12^\circ$, $+18^\circ$, $+24^\circ$, and dotted red for $+36^\circ$. The $\Lambda$CDM parameters are substantially stable for all these choices, except possibly for the gray curves. Notice how, with the model of eq.~(\ref{power_cutoff}), all $\Lambda$CDM parameters become more stable, even for very large masks.}
\label{fig:compared_LCDM}
\end{figure}

We begin by comparing the results for the six standard $\Lambda$CDM parameters, obtained with conventional or enlarged low--$\ell$ masks within $\Lambda$CDM and including $\Delta$.
The black and gray posteriors in fig.~\ref{fig:compared_LCDM} are for the $\Lambda$CDM spectrum of eq.~(\ref{tilt}), while the
blue and red ones are for the $\Lambda$CDM+$\Delta$ spectrum of eq.~(\ref{power_cutoff}).
All posteriors are nicely consistent and stable against Galactic masking, barring small but not insignificant shifts that occur for $\Lambda$CDM with a $+30^\circ$ extension (gray curves).
We interpret this behavior as a signature of the well known difficulty in reconciling high--$\ell$ and low--$\ell$ CMB likelihoods \cite{PlanckLike,Ade:2013kta}, which is exacerbated when using large Galactic masks \cite{Gruppuso:2013xba}.
On the other hand, the introduction of $\Delta$ stabilizes all $\Lambda$CDM parameters, even for very large masks.

\begin{figure}[ht]
\centering
\begin{tabular}{c}
\includegraphics[width=65mm]{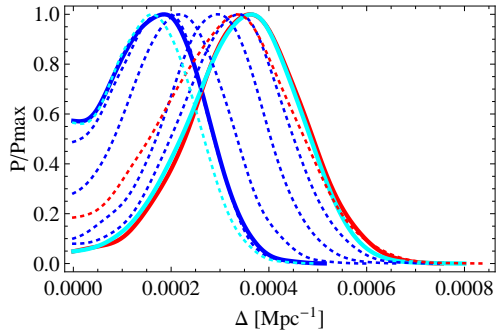}
\end{tabular}
\caption{\small Posteriors for the parameter $\Delta$ in the power law of eq.~(\ref{power_cutoff}), with color coding as in fig. \ref{fig:compared_LCDM}.
The cyan curves (dashed for the standard mask and solid for the $+30^\circ$ extension) take into account {\sc Planck} high $\ell$ polarization data (TT, TE, EE). Extending the Galactic mask results in a marked detection for $\Delta$ (see also Table \ref{percentages}).
The higher profile of the lower tail reflects the dependence on $\Delta^2$ of eqs. (\ref{power_cutoff}) and (\ref{coulomb_power}).}
\label{fig:compared_Delta}
\end{figure}

Posterior distributions for the additional parameter $\Delta$ are shown in fig.~\ref{fig:compared_Delta} for several Galactic masks, which now have a clear impact \cite{Gruppuso:2015zia}.
The detection levels for $\Delta$ are given in Table \ref{percentages}, along with estimated mean values and corresponding standard errors:
they increase monotonically with the masked area, from $88.5\%$ for $f_\mathrm{sky}=94\%$ up to $99.4\%$ for $f_\mathrm{sky}=39\%$.
However, decreasing $f_\mathrm{sky}$ further weakens the significance, due to increased sampling variance. The behavior of $\Delta$ is similar to the variance of the CMB pattern, which is known to decrease anomalously in extended Galactic masks \cite{Gruppuso:2013xba}.
Note that the inclusion of high $\ell$ polarization data (see the cyan curves in fig.~\ref{fig:compared_Delta}, dashed for standard mask and solid for $+30^\circ$ extension) does not modify significantly the constraints on $\Delta$.
This is expected since the modifications introduced in eq.~(\ref{power_cutoff}) impact only the large angular scales of the CMB anisotropies.

\begin{figure}[ht]
\centering
\begin{tabular}{c}
\includegraphics[width=65mm]{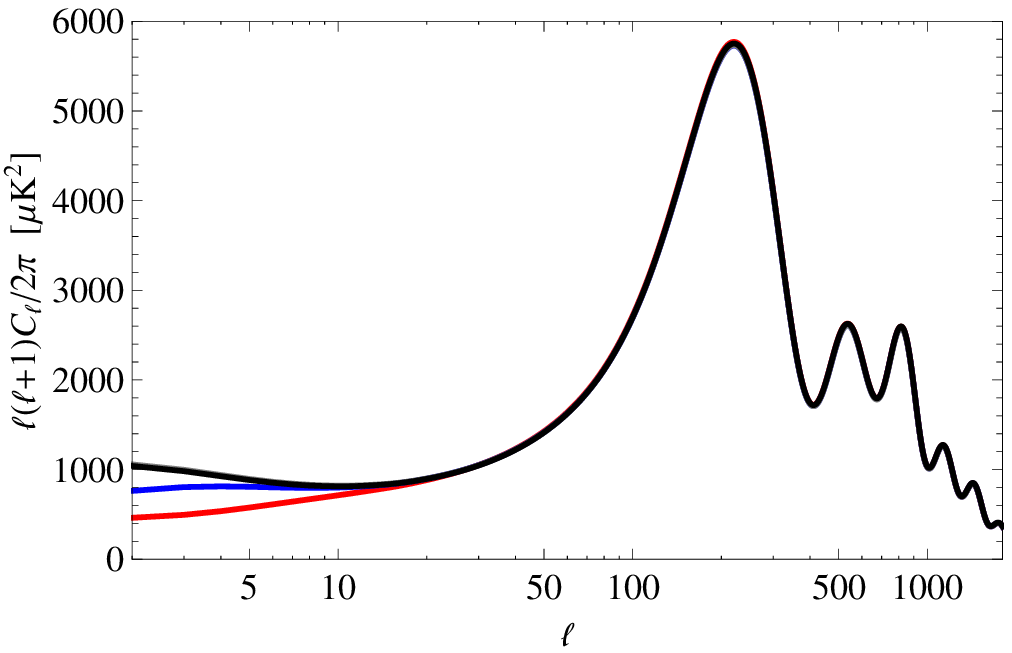} \\ \includegraphics[width=70mm]{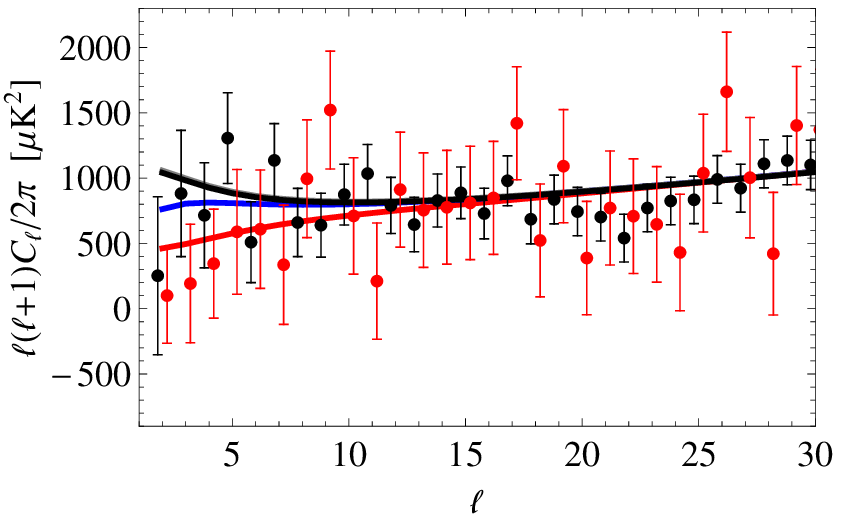}
\end{tabular}
\caption{\small {\it Upper panel}: best fit angular power spectrum models for $\Lambda$CDM with a standard mask (black), for $\Lambda$CDM with a $+30^\circ$ extension (gray, barely visible),
for $\Lambda$CDM$+\Delta$ with a standard mask (blue), and for $\Lambda$CDM$+\Delta$ with a $+30^\circ$ extension (red).
Only the $\ell \lesssim 15$ range is affected.
{\it Lower panel}: low-$\ell$ portion, with quadratic maximum likelihood estimates \cite{Gruppuso:2009ab} of the angular power spectrum and color coding as above.
Note how $\Lambda$CDM$+\Delta$ captures the decrease in power.}
\label{fig:fiducials}
\end{figure}

Fig.~\ref{fig:fiducials} displays the angular power spectrum coefficients of the fiducial models for four relevant cases. As expected, the standard $\Lambda$CDM model is modified for low $\ell$ by the additional parameter $\Delta$, and more evidently for an increased low--$\ell$ temperature mask: the large scale ``lack of power'' anomaly \cite{Copi:2013cya} disappears when $\Lambda$CDM$+\Delta$ is taken into account.
Is there, however, a statistically motivated reason for preferring $\Lambda$CDM+$\Delta$ to the standard $\Lambda$CDM? We have computed $\Delta \log {\cal L} = \log {\cal L}_{\Lambda CDM+\Delta} - \log {\cal L}_{\Lambda CDM}$, with ${\cal L}$ being the likelihood, at the best fit models,
finding $-0.3$ for $f_\mathrm{sky}=94\%$ and $-2.7$ for $f_\mathrm{sky}=39\%$.
The latter case yields a $98.0\%$ significance for the likelihood ratio test, see e.g. \cite{LikeRatioTest}. Alternatively, the Akaike Information Criterion yields the variations $+1.4$ for $f_\mathrm{sky}=94\%$ and $- 3.4$ for $f_\mathrm{sky}=39\%$, which points once more to the role of the Galactic mask. Our results for $\Delta$ are compatible with previous analyses made in \cite{destri} with different infrared cuts, but extended masks lead here to a higher significance.

\begin{table}
\centering
\caption{\small Mean value of $\Delta$ and standard deviation (second and third columns) and detection levels (fourth column) for all cases of fig.~\ref{fig:compared_Delta}. The parentheses identify cases in which {\sc Planck} high--$\ell$ polarization data (TE, EE) are also included. }
\label{percentages}
\begin{tabular}{cccc}
\hline
$f_\mathrm{sky} $ & Mean $\Delta$ & St.~Dev. $\Delta$ & Detection Level \\
$\%$ &  [$\times 10^4$ Mpc$^{-1}$] & [$\times 10^4$ Mpc$^{-1}$] & $\%$ \\
\hline
 94 & 1.7  (1.6) & 0.9 (0.8) & 88.5 (87.4) \\
 84 & 1.7  & 0.9 & 91.1 \\
71 & 2.1 & 1.0 & 94.7  \\
59 & 2.8 & 1.0 & 98.5 \\
49 & 3.2 & 1.1 & 98.9  \\
39 & 3.5 (3.4) & 1.1 (1.1) & 99.4 (99.4) \\
31 & 3.1 & 1.3 & 96.8 \\
\hline
\end{tabular}
\end{table}

\section{From $\Delta$ to a primordial scale}

For a Galactic mask extended by $30^\circ$, corresponding to $f_{sky}=39\%$, we found
\be
\Delta = (0.351 \pm 0.114) \times 10^{-3} \, \mbox{Mpc}^{-1}  \ ,
\label{Deltalargemask}
\ee
which differs from $0$ at $99.4\%$ C.L..
Let us now see why values of this type appear reasonable, in particular for the models of \cite{dkps,ks} that motivated this analysis. A typical feature is indeed that inflation lasts ${\cal O}(100)$ times the time scale set by $H_{\rm Infl}$, so that taking this fact into account and retracing the subsequent evolution of the Universe, one can convert eq.~(\ref{Deltalargemask}) into primordial length or energy scales at the onset of inflation,
\be
\Delta^{\rm Infl} \ = \ 3 \times 10^{14} \  e^{N-60} \times \sqrt{\frac{H_{\rm Infl}}{\mu_{Pl}}} \ \ {\rm GeV} \ ,
\label{delta0}
\ee
where $\mu_{Pl}$ is the reduced Planck mass.
The result depends on the number of $e$--folds, and demanding that $\Delta^{\rm Infl} \gtrsim H_{\rm Infl}$ yields the inequality
\be
e^{N-60} \ \gtrsim \ {8 \times 10^{3}} \ \sqrt{\frac{H_{\rm Infl}}{\mu_{Pl}}}\ .
\ee
For $H_{\rm Infl} \simeq 10^{14}\ {\rm GeV}$ this implies the reasonable bound $N \gtrsim 64$. Conversely, {\sc Planck} set the upper bound \cite{planck_XX}
\be
\frac{H_{\rm Infl}}{\mu_{Pl}} \ < \ 3.6 \times 10^{-5} \ ,
\ee
and making use of this result in eq.~(\ref{delta0}) yields
\be
\Delta^{\rm Infl} \ \lesssim \ 4 \times 10^{12} \ e^{N-60} \ {\rm GeV} \ ,
\ee
with an upper bound again ${\cal O}(10^{14})$ GeV for $N\simeq 64$.

\begin{figure}[ht]
\centering
\begin{tabular}{cc}
\includegraphics[width=38mm]{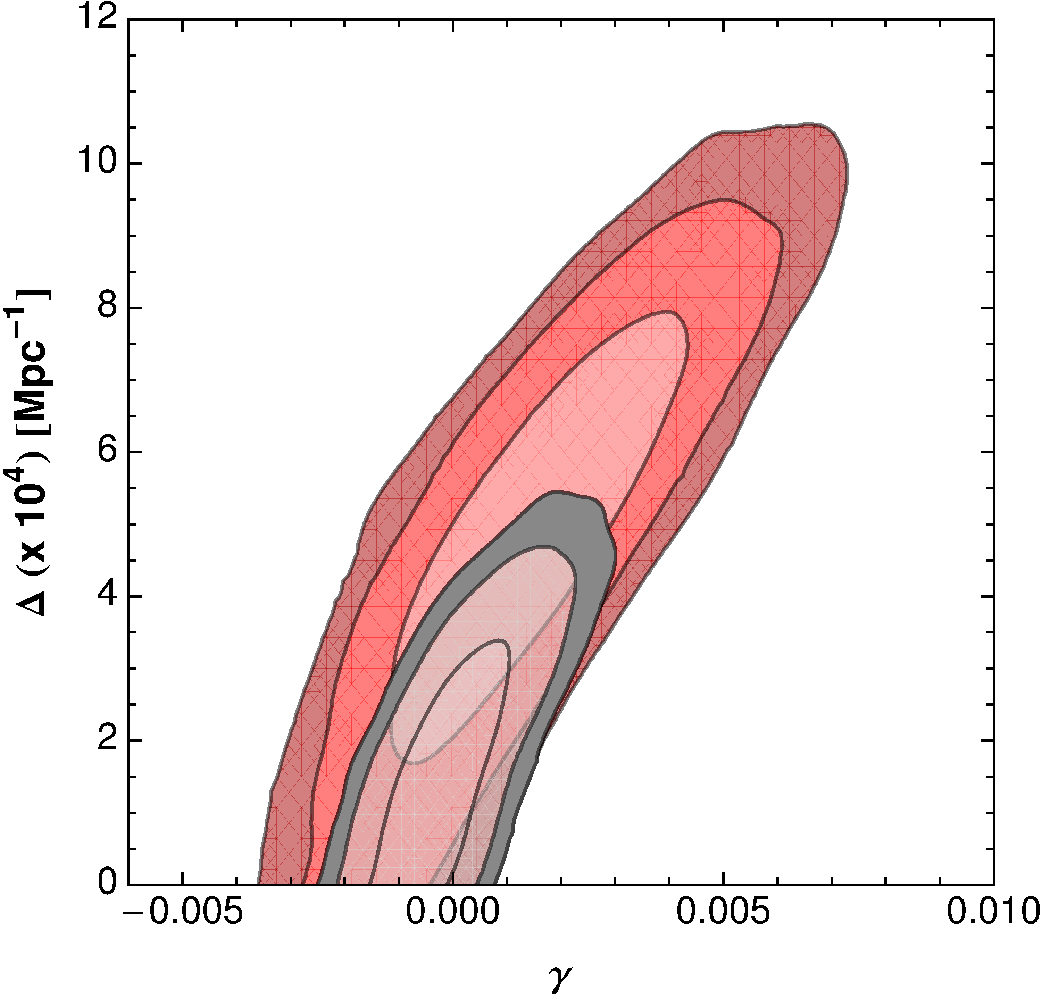} & \includegraphics[width=45mm]{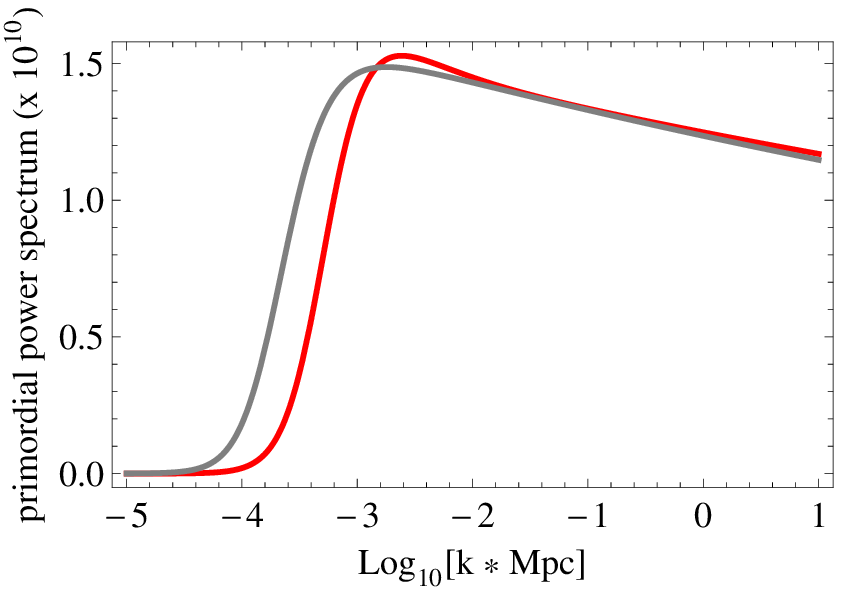}  \\
\end{tabular}
\caption{\small \emph{Left:} 2D contour plots for $\Delta$ and $\gamma$ of eq.~(\ref{coulomb_power}), for the standard mask (gray) and for the $+30^\circ$ mask (red). Introducing $\gamma$ does not affect the mean value of $\Delta$ but impacts its detection level, which lowers to 97.6\% (97.3\%) in the $+30^\circ$ mask for TT(+TE+EE)+lowP+lensing likelihood, an effect that can be partly ascribed to their correlation (fig.~\ref{fig:compared_spectra}).
\emph{Right:} primordial power spectra for the corresponding cases. A mild peak appears slightly preferred with the $+30^\circ$ mask.}
\label{fig:Deltagamma}
\end{figure}

\section{Conclusions}

The present epoch apparently confronts us with enticing clues on the future evolution of the Universe and remarkable windows on its past.
In this letter, we have provided some evidence that we might be observing, in the CMB, relics of the onset of inflation.
We have stressed how the lack of power observed at low $\ell$ is enhanced with a wider, and hence more conservative, Galactic mask. Introducing the infrared cutoff $\Delta$ of eq.\ (\ref{power_cutoff}), we were able to model this effect, detecting $\Delta = (0.351 \pm 0.114) \times 10^{-3} \, \mbox{Mpc}^{-1}$ at $99.4\%$ C.L.\ in a blindly chosen, but widely extended and aggressive Galactic mask, with $f_{sky}=39\%$.  As we have seen, this value would translate into primordial energy scales ${\cal O}(10^{14})$ GeV with about 64 $e$--folds of inflation. The resulting improvement in $\chi^2$ is consistent with the analysis in \cite{ks}: the features in fig.~\ref{fig:compared_spectra} are indeed \emph{stretched by wider Galactic masks}, with preferred low--$\ell$ angular power spectra that approach the model of eq.~(\ref{power_cutoff}) and become \emph{essentially independent of the initial condition} $\varphi_0$ \cite{as_erice}.
All in all, larger Galactic masks favor a detection of $\Delta$, because the CMB sky contains less low--$\ell$ power at high Galactic latitudes. This is an observational fact, derived under more conservative assumptions than those reflected in the standard mask of the {\sc Planck} likelihood,
but present data do not rule out the possibility that this behavior originate from a statistical fluke. Moreover, we have verified that the constraints on $\Delta$ are very stable when the high-$\ell$ polarization {\sc Planck} Likelihood (TE+EE) is included in the analysis (see fig.~\ref{fig:compared_Delta} and the Table).
It would be interesting to examine alternative sets of masks, since the improved detection of $\Delta$ might reflect the removal of known low--$\ell$ disturbances, for instance radio loops \cite{liumertsar}, or perhaps hint at others. The topic thus remains a high--priority target for future investigations of low--$\ell$ anomalies. Finally we have explored, albeit to a lesser extent, the second parameter $\gamma$ of eqs.~(\ref{coulomb_power}).
The preliminary analysis of fig.~\ref{fig:Deltagamma}, whose left panel displays the most relevant correlation, points toward slightly positive values of $\gamma$, compatibly with a mild pre--inflationary peak, while the detection of $\Delta$ becomes less significant.

Let us conclude by stressing that, if the CMB lack of power at low $\ell$ were due to a decelerating inflaton, it would be accompanied, in the same region, by an increased tensor-to-scalar ratio. This was noted in \cite{dkps} with reference to the class of models of \cite{bsb,climbing}, but the effect can be ascribed, in general, to the larger values attained by the slow--roll parameter $\epsilon$ in the relevant region.

\paragraph{Acknowledgments}
We are grateful to M.~Lattanzi, S.~Matarrese, L.~Pagano, and especially to B.~Partridge, for useful discussions and suggestions.
We acknowledge the use of computing facilities at NERSC (USA), of the HEALPix package \cite{gorski}, and of the {\sc Planck} Legacy Archive (PLA). Research supported by ASI through ASI/INAF Agreement I/072/09/0 for the Planck LFI Activity of Phase E2, and by INFN (I.S. Stefi, FlaG, InDark). NK was supported in part by the JSPS KAKENHI, Grant Number 26400253, while AS was supported in part by Scuola Normale. We would like to thank CERN Ph--Th, Scuola Normale Superiore, the Univ. of Ferrara and IASF--Bologna for their kind hospitality.

%
%
%

%
\end{document}